\documentclass[prl,twocolumn,showpacs,preprintnumbers,amsmath,amssymb]{revtex4}

\usepackage{graphicx}
\usepackage{dcolumn}
\usepackage{bm}

\begin{document}


\title{Bandwidth tuning triggers interplay of charge order and superconductivity
in two-dimensional organic materials}


\author{S. Kaiser$^{1}$,
M. Dressel$^{1}$, Y. Sun$^{1}$,
A. Greco$^{2}$,
J.A. Schlueter$^{3}$,
G.L. Gard$^{4}$,
and N. Drichko$^{1,5}$
}

\affiliation{%
$^{1}$ 1.~Physikalisches Institut, Universit\"at Stuttgart, Pfaffenwaldring 57, 70550 Stuttgart, Germany\\
$^{2}$ Facultad de Ciencias Exactas Ingenier\'ia y Agrimensura e Instituto de F\'isica Rosario (UNR-CONICET), Rosario, Argentina\\
$^{3}$ Material Science Division, Argonne National Laboratory, Argonne, Illinois 60439-4831, U.S.A\\
$^{4}$ Department of Chemistry, Portland State University, Portland, Oregon 7207-0751, U.S.A.\\
$^{5}$ Ioffe Physico-Technical Institute, St. Petersburg, Russia
}

\date{\today}

\begin{abstract}
We observe  charge-order fluctuations in the quasi-two-dimensional organic superconductor
$\beta^{\prime\prime}$-(BEDT-TTF)$_2$SF$_5$CH$_2$CF$_2$SO$_3$ both by means of vibrational spectroscopy, locally probing the  fluctuating charge order, and  investigating the in-plane dynamical response by infrared reflectance specctroscopy.   The decrease of effective electronic interaction in an isostructural metal suppresses both charge-order fluctuations and superconductivity, pointing on their interplay. We compare the results of our experiments with calculations on the extended Hubbard model.

\end{abstract}

\maketitle

From a naive point of view, superconducting and ordered insulating
states are incompatible. This is true, for example, for the competition of
charge-density wave and superconductivity \cite{Shelton86}, or stripes in cuprates \cite{Dumm02,Homes06}.
However, experimental and theoretical studies on materials with strong electronic correlations suggest that {\em fluctuations} of an ordered state may mediate superconductivity. Prime candidates for this mechanism are
magnetic order in heavy fermions \cite{Sato01,Levy} and high-temperature superconductors \cite{Sidis01}, incommensurate charge-density waves in dichalcogenides \cite{Dordevic03}, or fluctuating charge order in quasi-two-dimensional organic conductors \cite{Merino01,Dressel03,Dressel04,Powell06,Seo06,Merino07,Bangura05}.

In BEDT-TTF-based  1/4-filled conductors  the ground state can be tuned by modifying  effective electronic correlations via changing the bandwidth  \cite{Mori99,Seo04}(Fig.~1a): A  charge-ordered insulating state is observed when the effective  Coulomb repulsion  is large enough \cite{Yamamoto02,Takahashi,McKenzie01,Seo04,Wanatabe06}, while compounds with weaker effective electronic correlations are metallic \cite{Calandra02}. In the metallic state close to the metal-insulator phase boundary, charge fluctuations are observed \cite{Dressel03, Merino06,Takahashi}, while the response of coherent carriers is still present. These experimental results are in agreement to calculations on the extended Hubbard model. It is the minimum model that can describe a metal-insulator transition in quasi-two-dimensional molecular conductors with 1/4-filled conduction band [7]. It takes into account
the effective on-site $U/t$ and inter-site $V/t$ Coulomb repulsion, where $t$ is the hopping integral related to the bandwidth.
This model  predicts that fluctuations of checker-board charge order (CO) can act as an attractive interaction of quasiparticles forming Cooper pairs and lead to a superconducting state \cite{Merino01,Merino03}. In this Letter we  present an experimental evidence for a bandwidth tuned  CO  fluctuations  in the normal state of $\beta^{\prime\prime}$ family of quasi-2D organic conductors. We see an unambiguous {\rm relation} between the  presence of CO fluctuations and  superconductivity and discuss its origin.

\begin{figure}
\includegraphics[width=7cm]{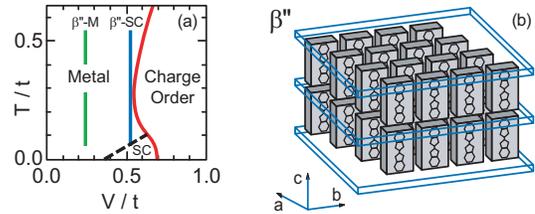}
\caption{\label{fig:1}(Color online)(a) Phase diagram for the quarter-filled organic conductors calculated by extended Hubbard model. The dashed line for superconductivity is put schematically based on Ref.[\onlinecite{Merino01}]. Colored lines show approximate position of the studied compounds in the phase diagram. (b) A schematic view of a 3D structure of the layered BEDT-TTF based crystals, the rectangulars mark BEDT-TTF molecules.}
\end{figure}

The studied materials are  layered compounds, where a slightly anisotropic quasi-two-dimensional conducting electronic system of the $(ab)$ plane is created  by the overlap of the neighboring BEDT-TTF (bis-(ethylenedithio)tetrathiafulvalene) molecules~\cite{Wang99} (Fig.~1b). The bandwidth is tuned by changing the size of the anion, by so called 'chemical pressure'.
$\beta^{\prime\prime}$-(BEDT-TTF)$_2$\-SF$_5$\-CH$_2$\-CF$_2$\-SO$_3$ (named as $\beta^{\prime\prime}$-SC later on) \cite{Geiser96}  becomes superconducting at $T_c\approx 5$~K [inset of Fig.~\ref{fig:1}(a)] and  was suggested to be an ideal model to investigate the
interplay between  CO and superconductivity in a 1/4-filled system \cite{Merino01}. In the isostructural compound $\beta^{\prime\prime}$-(BEDT-TTF)$_2$SO$_3$CHFSF$_5$ (named as $\beta^{\prime\prime}$-M) the effective Coulomb repulsion between electrons is reduced by chemical pressure  \cite{Schlueter01,Wang99}, and the compound is metallic.

The molecular structure
of conducting layers of BEDT-TTF-salts gives us a unique opportunity to locally
probe the charge distribution between the lattice sites by infra-red (IR) reflectance. We follow a temperature dependence of the sensitive to site charge B$_{1u}(\nu_{27})$ vibrational  mode of BEDT-TTF molecules, observed in the out-of-plane IR spectra  \cite{Yamamoto05,Dressel04}.
This temperature dependence compared to the in-plane temperature-dependent  d.c. conductivity gives a spectacular result presented in Fig. \ref{fig:1}. At room temperature both compounds show a wide single line of   B$_{1u}(\nu_{27})$ vibration. The band stays wide but single for the $\beta^{\prime\prime}$-M in the whole temperature range. When the effective correlations increase  in the case of  $\beta^{\prime\prime}$-SC, the spectra show a splitting  $\Delta
\nu=22~{\rm cm}^{-1}$ of this charge sensitive vibration at temperatures below 200 K. No changes in any other vibrational feature occur and our x-ray studies at $T= 300$, 123 and 20 K   evidence
that the symmetry of the unit cell of $\beta^{\prime\prime}$-SC does not change
with temperature. The observed  splitting suggests that the system is tuned into a state with a charge disproportionantion
$\Delta\rho \approx 0.2e$ between neighboring lattice sites \cite{Yamamoto05}. Nevertheless the compound stays metallic with only a slight change in the slope of d.c. resistivity temperature dependence below 200 K \cite{Glied08}. We interpret the result as  a formation of a weak fluctuating  CO. The analysis of the width of the vibrational bands suggests a characteristic time of charge  fluctuations between the sites not less than $10^{-12} s$ \cite{Gutowsky53}. $\beta^{\prime\prime}$-SC with  fluctuating CO  shows  superconductivity with $T_c$=5 K, while $\beta^{\prime\prime}$-M with $\Delta\rho$  below 0.1$e$ stays metallic.

To study the response of the quasi-two-dimensional conducting system, we perform  IR reflectance measurements in the  $ab$-plane plane of high-quality single crystals ($3\times 1.2 \times 0.15~{\rm mm}^3$)  in $8-8000~{\rm cm}^{-1}$ frequency  and $300-1.8$~K temperature ranges by using FTIR and THz spectrometers. In Fig. 3 and 4 a,b we show the conductivity spectra obtained by a Kramers-Kronig transformation, using standard extrapolations\cite{DresselGruner02}.
\begin{figure}
\includegraphics[width=9cm]{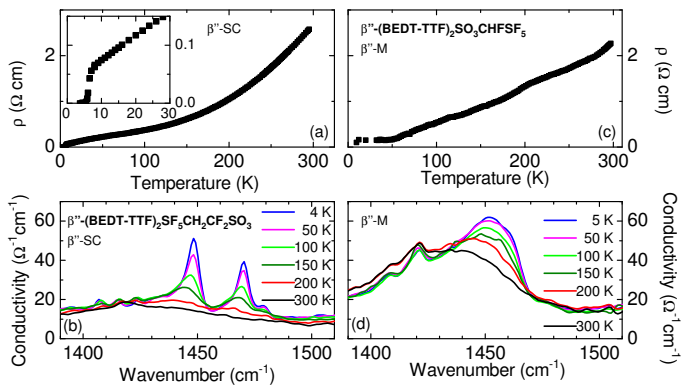}
\caption{\label{fig:1}(Color online) Temperature-dependent dc
resistivity in the conducting plane of (a) $\beta^{\prime\prime}$-SC  and (c) $\beta^{\prime\prime}$-M. Optical conductivity perpendicular to the conducting layers in the region of B$_{1u}(\nu_{27})$ BEDT-TTF vibration: (b) $\beta^{\prime\prime}$-SC: a band of B$_{1u}(\nu_{27})$ splits into two components  at 1448 and 1470~cm$^{-1}$ on temperature decrease; (c) $\beta^{\prime\prime}$-M: B$_{1u}(\nu_{27})$ vibration of BEDT-TTF is a single wide band down to 10 K. }
\end{figure}
 These measurements cover the energy range of the conductance band formed by the overlap of the BEDT-TTF molecular orbitals~\cite{Dressel04}. For a metal we expect all the respectful spectral weight to concentrate in a zero-frequency (Drude) peak of the conductivity spectrum.  As in case of many other materials with competing interactions \cite{BasovRMP05,DresselGruner02}, even in the metallic state the spectra do not contain only a Drude component (see Fig. 3). For both compounds at room temperature   most of the spectral weight is found in the  wide  features in mid-infrared (MIR) with maxima at about 1000 and 2500 cm$^{-1}$. On cooling, the spectral weight shifts to low frequencies, and a Drude peak appears, as is  typical for organic conductors \cite{Takenaka05,Drichko06,Faltermeier07}.

The Drude spectral weight of $\beta^{\prime\prime}$-M  does not change below 150 K, respectively the  Drude plasma frequency $\Omega_p$ stays constant (see Fig.~4c). In contrast to that, the Drude spectral weight of more correlated
 $\beta^{\prime\prime}$-SC grows on further cooling on the expense of the band at 2500~cm$^{-1}$. This re-entrant behavior is suggested for a metal close to  CO by the calculations on the extended Hubbard model: the phase border between CO and metal shifts to higher values of $V$ at low temperatures, see Fig.~1a. The estimated temperature of the re-entrant behavior is $T=0.2t$, that gives us 140 K with $t$=0.06eV that is a typical value for these compounds.

At 10~K for the $\beta^{\prime\prime}$-M we observe a well-defined Drude-like peak below $\approx$ 200 cm$^{-1}$, a band at  $\approx$ 700 cm$^{-1}$ shifted down from 1000 cm$^{-1}$ and a wide band at 2500 cm$^{-1}$ (see the insert in Fig.3b). A weak band at about 300 cm$^{-1}$ might be already distinguished in the $\beta^{\prime\prime}$-M spectra. It becomes very intense for   $\beta^{\prime\prime}$-SC and dominates the low-frequency response, while the Drude-like peak is extremely narrow and  contains about  5 \% of the spectral weight (see the insert in Fig.3a). The 300 cm$^{-1}$ band in the spectra of $\beta^{\prime\prime}$-SC  starts to increase at temperatures when the  charge disproportionation sets in.

The interpretation of experimental conductivity spectra for 1/4-filled organic conductors was based on the results of numerical solutions of the extended Hubbard model on a square lattice\cite{Merino03,Dressel03,Drichko06}.   In a  metallic state close to the charge-order, where charge fluctuations are important, this theory expects three features in conductivity spectra: (i)~A broad band in MIR range which we associate with site-to-site transitions within fluctuating CO pattern. This is a transition between  ``Hubbard-like'' bands (see Fig. 8 and 9 in Ref. [\onlinecite{Merino03}]) (ii)~a Drude peak describing the coherent-carriers response, due to the narrow peak of DOS at Fermi energy, and (iii)~a finite-frequency `charge fluctuation band' due to transitions between the DOS at Fermi energy and the Hubbard-like bands, observed for 1/4-filled organic conductors typically at 700-500~cm$^{-1}$ \cite{Drichko06,Merino06}. The presence of the latter feature in the spectra proves that we observe  CO fluctuations, and not a separation between a metallic and charge ordered components. All these three features can be well distinguished in the spectra of $\beta^{\prime\prime}$-M and $\beta^{\prime\prime}$-SC (Fig.3).
\begin{figure}
\includegraphics[width=\columnwidth]{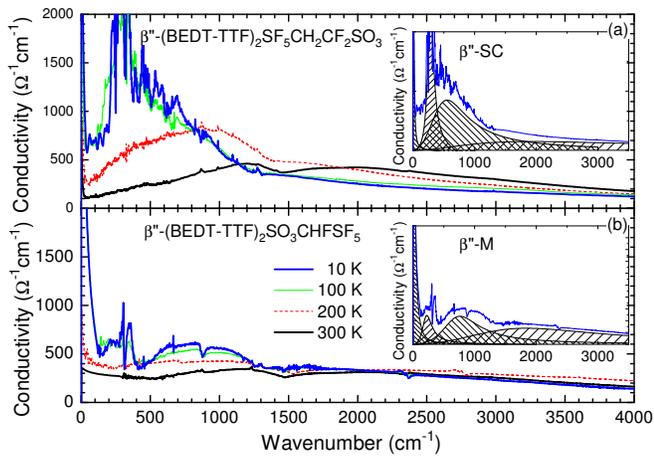}
\caption{\label{fig:2}(Color online)
In-plane conductivity ($E \parallel b$) of (a) $\beta^{\prime\prime}$-SC  and (b) $\beta^{\prime\prime}$-M at different temperatures. On the inserts we show conductivity spectra at 10 K with the results of the Drude-Lorentz fit and marked electronic bands, discussed in the text. The spectra in the perpendicular direction show the same electronic features.
\label{Fig2}}
\end{figure}

The important and  striking result of this work is an additional feature that appears at about 300~cm$^{-1}$ and increases in intensity  on getting close to CO, when  going from $\beta^{\prime\prime}$-M to $\beta^{\prime\prime}$-SC. In the metallic state  with fluctuating CO in  $\beta^{\prime\prime}$-SC the Drude peak gets extremely narrow and the band at 300~cm$^{-1}$ very intense, while their spectral weights simultaneously increase on cooling.
This suggests an assignment of the band to the scattering of charge carriers on CO fluctuations near the CO  phase. In respective theoretical picture, close to  CO and in the re-entrant regime at $T< 0.2t$ (Fig.1a)  collective charge fluctuations are very soft \cite{Merino06,Merino03}. The interaction between charge carriers and the respective bosonic low-energy plasmon modes leads to self-energy effects and electronic incoherent structures which manifest as a peak in the conductivity at $\omega\sim 0.6t$.  However,  the extremely high intensity
\begin{figure}
\includegraphics[width=8cm]{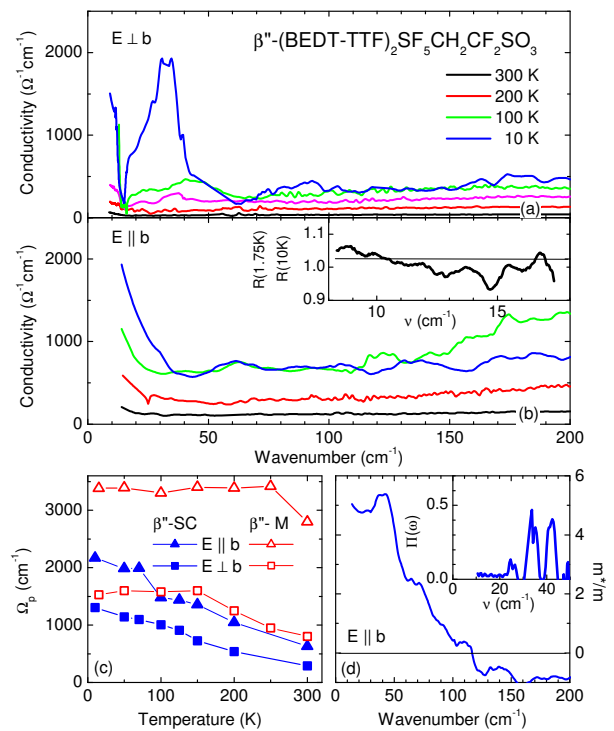}
\caption{\label{fig:3} (Color online)
(a)~The conductivity in $E \perp b$ of
$\beta^{\prime\prime}$-SC shows a narrow Drude contribution and the lattice mode at about 40 cm$^{-1}$ coupled to charge carriers. (b)~The conductivity in $E \parallel b$ direction; the insert shows the opening of the superconducting gap as the ratio  R(1.75 K)/R(10 K). (c) Temperature dependence of Drude plasma frequency. Note the re-entrant behavior.(d)~the effective mass at 10 K calculated by the extended Drude analysis, the insert shows $\bar{\Pi}(\omega)$ at the lowest frequencies.  }
\end{figure}
of this feature at 300 cm$^{-1}$ in $\beta^{\prime\prime}$-SC spectra is not reproduced by the calculations on the extended Hubbard model of Ref.[\onlinecite{Merino03}].

An  alternative theoretical approach proposes a description of the CO state that includes  coupling to the lattice, resulting in a polaronic state \cite{Ciuchi08}.
This model deals with an ordering transition and charge disproportionation $\Delta\rho$ of above 0.5$e$, and not a metallic state. Important though, as consequence of a strong coupling to the lattice a CO electronic feature is extremely enhanced and a pseudogap present in the metallic state.
In accord to this work, we suggest that  polaronic effects enhance the low-energy band, and to fully describe the results a coupling to the lattice should be taken into account in some more sophisticated model.
This increasing importance of coupling to the lattice in $\beta^{\prime\prime}$-SC can be explained by a large renormalization of the electron-phonon coupling due to CO fluctuations close to CO \cite{foussats05}.

The electron-phonon coupling is a reason of an increased  intensity of a  phonon mode at $30-40~{\rm cm}^{-1}$  on cooling.
 This mode is observed with $E \perp b$ (Fig.~\ref{fig:3} a) as charge disproportionation appears  below 200~K. The lattice modes at these frequencies are vibrations that modulate the distance
between the sites \cite{Girlando00}. A mode involving a vibration of two  neighboring lattice sites (BEDT-TTF molecules) will appear or increase its intensity in a CO state. While $\Delta \rho\approx 0.2e$ remains constant, the intensity of the
40~cm$^{-1}$ mode increases upon cooling together with the low-frequency electronic features, showing the strong
coupling to the charge-carrier response. This agrees with an expected enhancement of  the electron-phonon coupling near CO \cite{foussats05}.

 We obtain further information on electron-bosonic coupling from the extended Drude analysis and optical memory functions \cite{BasovRMP05, BasovRMP10} for $E \parallel b$, where the zero-frequency peak in conductivity is wide enough to perform this analysis.
 An extended Drude model reveals the frequency-dependence of the scattering rate and effective mass $\frac{m^*(\omega)}{m_b}=\frac{\Omega_p^2}{4
\pi}\frac{\sigma_2}{\omega \left| \sigma \right|^2}$, the latter shown in  Fig.~\ref{fig:3}d, where the plasma frequency  $\Omega_p$ is received from the spectral weight of the Drude-like peak.
The values of the effective mass are in good agreement with the value of  3.9$m_e$ received from
specific heat measurements \cite{Wanka98}, while higher than  1.9$m_e$ received from Shubnikov-de-Haas oscillations.
Further, from our data we get an approximate values of $\bar{\Pi}(\omega)$\cite{note}, the coupling of electrons to bosonic fluctuations:  $\bar{\Pi}(\omega) \approx \frac{1}{2\pi}\frac{d^2}{d\omega^2} [ \omega \frac{1}{\tau(\omega)}]$ (see \cite{BasovRMP10} and references therein) (insert in Fig.~\ref{fig:3}d) . The  maxima in $\bar{\Pi}(\omega)$  around 40~cm$^{-1}$ suggest the approximate frequencies of the coupled bosonic excitations, while the peak in conductivity in $E \perp b$ directly shows the coupling of charge carriers to the IR active 40~cm$^{-1}$ feature.  The phonons at about 40~cm$^{-1}$  also give an important contribution
to the specific heat of
$\beta^{\prime\prime}$-SC \cite{Wanka98}.
%

Below $T_c=5.4$~K we observe the opening of a superconducting gap (see insert in Fig.4 b), where frequencies below $2\Delta\approx 12$~cm$^{-1}$ the reflectivity jumps to 1 at $T=1.8$~K. Important, we do not see any change in the charge disproportionation features while observed the superconducting gap in optics.
The size of the superconducting gap $2\Delta(0) \approx
12$~cm$^{-1}$ suggests weak coupling with $2\Delta/k_B T_c \approx
3.3 $.

In conclusion, in this letter we present a study of a bandwidth tuning of  fluctuating charge order in  quasi-two-dimensional  quarter-filled organic conductors. We directly estimate the charge disproportionation between the lattice sites by vibrational spectroscopy, as well as detect spectra of the quasi-two-dimensional electronic system characteristic of charge order fluctuations.   We show that on the increase of effective correlations,  both  CO fluctuations and superconductivity appear in  $\beta^{\prime\prime}$-(BEDT-TTF)$_2$SF$_5$CH$_2$CF$_2$SO$_3$. The strong charge fluctuations affect the density of states and the Fermi surface. That explains  a non-expected  discrepancy found between the calculated and measured Fermi surface for  $\beta^{\prime\prime}$-(BEDT-TTF)$_2$SF$_5$CH$_2$CF$_2$SO$_3$ \cite{Wosnitza07}.

The  MIR spectra of BEDT-TTF 1/4-filled salts: the $\beta^{\prime\prime}$ presented in this work, $\theta$-phases (for example \cite{Tajima00,Yamamoto02,Wang01}), and  $\alpha$-phases \cite{Drichko06} are  amazingly similar and are all well described by  the extended Hubbard model  (Ref. \cite{Calandra02}) that takes into account only electronic correlations. However, this model cannot describe the high intensity of the band due to scattering of charge carriers on CO fluctuations for $\beta^{\prime\prime}$-SC, we suggest that here the  interaction with the lattice should be taken into account.

Concerning the origin of superconductivity, the enhancement of electron-phonon coupling by CO fluctuations would  also  enhance superconducting pairing\cite{foussats05} near CO  with respect to the pure electronic case\cite{Merino01}. This would put our results in agreement with previous calculations showing the importance of both electron-phonon and electron-molecular vibrations coupling for superconductivity \cite{Girlando02}, but also explain the decisive role of charge order fluctuations, proved by our work.

We thank S. Yasin,  A. Dengl, and M. Herbik
for performing measurements and J. Merino, M. Dumm, Ch. Hotta, U. Nagel, T. R\~o\~om and N.P. Armitage for valuable discussions. The work was supported by DFG. ND acknowledges the support by the
Margarethe von Wrangell program. Work at Argonne National
Laboratory is sponsored by US DOE
contract No. DE-AC02-06CH11357. Work at Portland was supported by
NSF (Che-9904316).


\begin{thebibliography}{99}
\bibitem{note}Following \protect\cite{BasovRMP10} we use the notation $\bar{\Pi}(\omega)$ as an analogy of the Eliashberg function, that includes a coupling not only to phonons, but to any kind of bosonic exitations, including CO fluctuations.
\end{thebibliography}
\end{document}